\documentclass[%
superscriptaddress,
]{revtex4-2}

\usepackage{physics}
\usepackage[version=4]{mhchem}
\usepackage{siunitx}
\usepackage{graphicx}
\usepackage{dcolumn}
\usepackage{bm}

\begin{document}


\title {A Self-Adaptive First-Principles Approach for Magnetic Excited States}

\author{Zefeng Cai}
\affiliation{%
 School of Materials Science and Engineering, Tsinghua University, Beijing 100084, People’s Republic of China}%

\author{Ke Wang}
\affiliation{%
 School of Materials Science and Engineering, Tsinghua University, Beijing 100084, People’s Republic of China}%


\author{Yong Xu}
\affiliation{
 Department of Physics, Tsinghua University, Beijing 100084, People’s Republic of China
}%

\author{Su-Huai Wei}
\affiliation{
 Beijing Computational Science Research Center, Beijing 100193, People’s Republic of China
}%

\author{Ben Xu}
\thanks{Corresponding author}
\email{bxu@gscaep.ac.cn}
\affiliation{
 Graduate School of China Academy of Engineering Physics, Beijing 100088, People’s Republic of China
}%




\date{\today}

\footnotetext{This is the text of the footnote.}

\begin{abstract}
The profound impact of excited magnetic states on the intricate interplay between electron and lattice behaviors in magnetic materials is a topic of great interest. Unfortunately, despite the significant strides that have been made in first-principles methods, accurately tracking these phenomena remains a challenging and elusive task. The crux of the challenge that lies before us is centered on the intricate task of characterizing the magnetic configuration of an excited state, utilizing a first-principle approach that is firmly rooted in the ground state of the system. We propose a versatile self-adaptive spin-constrained density functional theory formalism. By iteratively optimizing the constraining field alongside the electron wave function during energy minimization, we are able to obtain an accurate potential energy surface that captures the longitudinal and transverse variations of magnetization in itinerant ferromagnetic Fe. Moreover, this technique allows us to identify the subtle coupling between magnetic moments and other degrees of freedom by tracking energy variation, providing new insights into the intricate interplay between magnetic interactions, electronic band structure, and phonon dispersion curves in single-layered CrI$_3$. This new methodology represents a significant breakthrough in our ability to probe the complex and multifaceted properties of magnetic systems. 
\end{abstract}

\maketitle

Excited magnetic states possess longitudinal and transverse perturbations of magnetic moments that deviate from ground-state magnetic configurations. They are crucial for understanding magnetic ordering, magnetic phase transitions, and other magnetic phenomena. Further, magnetic excited states are prevalent in frustrated magnets \cite{LeeNat2002,Pelissetto_PR2002, Balents_nat_2010,nisoli_RMP2013}, multiferroics \cite{Tokura_RPP,XiangHJ2008,Fabregesprl2009,Lss_PRB_2005}, superconductors \cite{PWAnderson_PRB_1984, Sidis_PRL_1999, Kuwabara_PRL_2000, Mazin_PRL_1997, Tranquada_PRL_1990,Monthoux_PRL_1992, Mook1998,Lumsden2010, Chen2019,Toru_AP_2000, Moriya_2003}, topological magnets \cite{xuyong_SA_2019}, etc, and are of crucial importance to phenomena such as superconductivity \cite{PWAnderson_PRB_1984, Sidis_PRL_1999, Kuwabara_PRL_2000, Mazin_PRL_1997, Tranquada_PRL_1990,Monthoux_PRL_1992, Mook1998,Lumsden2010, Chen2019,Toru_AP_2000, Moriya_2003}, quantum critical point \cite{Rosch_PRL_1999,George_PRB_2001,Ishida_PRL_2002, Hilbert_RMP_2007}, quantum phase transition \cite{jszhang_Science_2013}, and the quantum anomalous Hall effect \cite{LYH_PRL_2021}. The above exotic phenomena stems from the delicate interaction between spin and other degree of freedoms (DoF) such as charge, orbital, and lattice. The elucidation of the microscopic mechanism of these coupling effects need to evaluate the second or higher order derivatives of the energy with respect to the magnetic moment and other DoFs. In order to do so, the magnetic moment must be considered as a variable that can be altered independently from the other DoFs. And the manifestation of its variation, i.e. the excited magnetic states, is hence critical to understand the mutual interactions, and is significantly challenging to simulate despite its ubiquity. 

However, the commonly used density functional theory (DFT) \cite{kohn1965self} was originally developed to describe ground states. In their seminal works, Dederichs \textit{et al.}\cite{dederichs1984ground} and Dudarev \textit{et al.}\cite{ma2015constrained,Dudarev_PRM_2019} introduced a Lagrange multiplier $\lambda$ to constrain the magnetic moments when solving the Kohn-Sham equation as $\nabla_{\phi_i} L=0$ \cite{Richter_JMMM1995, dederichs1984ground, G.M.Stock_PMB_1998, wu2005direct,kurz2004ab, ma2015constrained}. However, it is difficult to simultaneously reach the global minimum of Lagrangian and meet the constraint, and thus it is not guaranteed that the magnetic moment $\mathbf{M}$ gets to its constrained target. This situation is particularly severe in itinerant metallic magnetic materials \cite{zimmermann_PRB_2019}. Therefore, the magnetic moment is difficult to control and to be considered as a completely independent DoF, and the energy of corresponding system is difficult to identify. These uncertainties escalate when the energy is differentiated by an infinitesimal variation of magnetic moment or other DoF, to evaluate the coupling strength, such as the magnetoelectrical coupling constant. 

Another unfavorable fact is that these methods require the user to be highly proficient at interactively adjusting the optimization strategy \cite{kurz2004ab, Dudarev_PRM_2019}, which makes these methods highly prone to numerical error. This is because that the optimization is conducted only in the electronic energy functional minimization, where the constraining parameter $\lambda$ is set as a pre-defined constant. $\lambda$, when improperly chosen, leads the system to a local minimum ($\lambda$ too small) or to divergence ($\lambda$ too big). In additional, this fixed $\lambda$ is difficult to adapt to actual spin fluctuations where different magnetic components exhibit different deviations from the original ground state, which also makes the Hamiltonian numerically unstable to be diagonalized. Recently, Hegde \textit{et al.}\cite{hegde2020atomic} developed a self-consistent method to impose spin constraints and sped up the convergence significantly. However, this method only deals with collinear magnetic orders, which hinder the study of complex noncollinear spin fluctuations. 

To address these deficiencies, in this letter, we proposed a self-adaptive spin-constrained DFT scheme, where Kohn-Sham orbitals and the constraining vectors are updated iteratively to reach the global optimum of the Lagrangian and the target magnetic moments. In this scheme, the on-site adaptable constraint $\left\{\bm{\lambda}_I\right\}$ are imposed in the form of a local vector field. The amplitude and orientation of the constraining vector field $\left\{\bm{\lambda}_I\right\}$ depend on the local magnetic environment of atom $I$ and vary from component to component. Equipped with this strategy, the magnetic moment can be altered independently from the other DoFs in our non-collinear first-principles scheme. Our approach now paves the way for the investigation of spin fluctuation and its influence on other DoF. To exemplify this method, we then apply it to two scenarios. The first is a spin configuration with confined rotation, using ferromagnetic materials Fe and layered magnetic materials \ce{CrI3} as examples. To simulate real-world fluctuations, it is later generalized to study the electronic band structure of \ce{CrI3} with random magnetic orientations. 


\begin{figure}[tb]
\includegraphics[width=0.8\linewidth, draft=false]{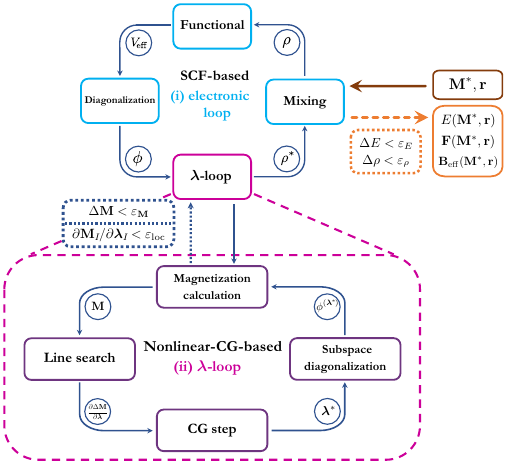}
\caption{\label{fig:algo} Scheme of self-adaptive spin-constrained DFT method \textit{DeltaSpin}. The outer electronic loop updates electron density $\rho$, while inner $\lambda$-loop updates constraining multiplier $\bm{\lambda}$. (Subscripts for indexing atoms and orbitals are omitted for clarity.)}
\end{figure}

\vspace{5mm}

\noindent {\large\textbf{Results}}

\noindent \textbf{Self-adaptive spin-constraining algorithm.} Our scheme is sketched in Fig.\,\ref{fig:algo}. It is based on the optimization of the Lagrangian function $L$ under the constraint that the magnetic moment of atom $I$, that is, $\mathbf{M}_I\left(\left\{\phi_i\right\}\right)$, is the same as the target value denoted by $\mathbf{M}_I^*$. To implement this, we design two nested loops updating the Kohn-Sham orbitals $\left\{\phi_i\right\}$ and the Lagrangian multiplier $\left\{\bm{\lambda}_I\right\}$ respectively.

(i) \textit{Electronic loop}: For a particular $\bm{\lambda}_I$, we can solve 
\begin{equation}
    \frac{\delta E_{\mathrm{KS}}}{\delta \bra{\phi_i}}-\sum_I \bm{\lambda}_I \cdot \frac{\delta \mathbf{M}_I}{\delta \bra{\phi_i}} =\epsilon_i\ket{\phi_i}\label{eq:ks}
\end{equation}
to obtain a self-consistent solution $\left\{\phi_i\left(\bm{\lambda}_I\right)\right\}$ and the magnetic moment $\mathbf{M}\left(\left\{\phi_i\left(\bm{\lambda}_I\right)\right\}\right)$. The atomic magnetic moment $\mathbf{M}_I$ is generally defined as $\mathbf{M}_I\left(\left\{\phi_i\right\}\right) = \mathrm{Tr}({\rho}\,{\sigma}\hat{\mathcal{W}}_I)$, where $\rho$ is the density matrix, ${\sigma}$ is the Pauli matrix, $\hat{\mathcal{W}}_I$ is a pre-defined weight operator for atom $I$. Thus, Eq.\,\ref{eq:ks} is obtained as
\begin{equation}
    \hat{\mathcal{H}}_{\mathrm{KS}}\ket{\phi_i}
    -\sum_{I}\bm{\lambda}_I\cdot\mathrm{\sigma}\hat{\mathcal{W}}_I \ket{\phi_i}=\epsilon_i\ket{\phi_i}. \label{eq:ks_m}
\end{equation}
Note that $\bm{\lambda}_I$ acts as a constraining field \cite{dederichs1984ground}. 

(ii) \textit{$\lambda$-loop}:  We can progressively update $\left\{\bm{\lambda}_I\right\}$ by optimizing
\begin{equation}
    \min _{\bm{\lambda}_I}\left|\mathbf{M}\left(\left\{\phi_i\left(\bm{\lambda}_I\right)\right\}\right)-\mathbf{M}^*\right|^2,
\end{equation}
that is, the error $\Delta \mathbf{M}$ between the current magnetization and the target, using the nonlinear conjugate gradient (NCG) method. For an updated $\bm{\lambda}_I^{*}$, we can solve Eq.\,\ref{eq:ks_m} via diagonalization in the subspace spanned by the old orbitals $\left\{\phi_i\left(\bm{\lambda}_I\right)\right\}$ to obtain a new set of orbitals $\left\{\phi_i\left(\bm{\lambda}_I^{*}\right)\right\}$, and thus, the moment $\mathbf{M}(\bm{\lambda}_I^*)$. 
In this way, the optimization can continue until a minimum is achieved. 
A perturbation-like scheme similar to that in Ref.\,\cite{hegde2020atomic} was chosen, which significantly reduced the computational cost compared to diagonalization in the entire basis set. In addition, the $\lambda$-loop converged rapidly because the function $\mathbf{M}\left(\left\{\phi_i\left(\bm{\lambda}_I\right)\right\}\right) = \partial^2 L/{\partial\bm{\lambda}_I^2}$ is proved to be almost monotonous according to the first-order perturbation theory ($L\left(\phi_i, \bm{\lambda}_I, \epsilon_i\right)$ is a concave function of $\bm{\lambda}_I$ at $0\,\mathrm{K}$ \cite{wu2005direct}). 


Certain details of the scheme deserve further discussion. In the $\lambda$-loop, the gradient of the object function can be approximated as
$
 2\,(\mathbf{M}_I - \mathbf{M}_I^{*}) \frac{\partial \mathbf{M}_I}{\partial \bm{\lambda}_I},
$
based on the assumption that $\pdv*{\mathbf{M}_{I^\prime \neq I}}{\bm{\lambda}_I}$, which is the non-local response to the on-site constraining field, is negligible. 
${\partial \mathbf{M}_I}/{\partial \bm{\lambda}_I}$ can be then obtained by lower-complexity estimation strategies for the NCG iteration. To provide better convergence, 
we control both $\Delta \mathbf{M} = \sqrt{|\mathbf{M}\left(\left\{\phi_i\left(\bm{\lambda}_I\right)\right\}\right)-\mathbf{M}^*|^2}$ and the main diagonal of the gradients $\pdv*{\mathbf{M}_{I}}{\bm{\lambda}_I}$ in the $\lambda$-loop. We first introduce a gradually tightened criterion $\varepsilon_{\mathbf{M}}$ for the former, preventing the constraining field from reaching an early stage local minimum. In case of the latter, the loop is programmed to stop when the local response was smaller than an empirical value $\varepsilon_\mathrm{loc}$ ($\varepsilon_\mathrm{loc} \approx 1\,\mu_B^2/\mathrm{eV}$ works well across our limited tests). This criterion, which is based on the aforementioned localization assumption, is of particular importance. Additionally, the weight operator in Eq.\,\ref{eq:ks_m} is often an integral in a real-space ball of radius $r_{\mathrm{cut}}$ with a smoothed boundary as $\hat{\mathcal{W}}_I = \int\mathrm{d}\mathbf{r}\,f(r_{\mathrm{cut}}-\left|\mathbf{r}-\mathbf{r}_I\right|)\ket{\mathbf{r}}\bra{\mathbf{r}}$. It may take a different form such as the Mulliken partitioning in atomic orbital basis DFT \cite{cuadrado2018implementation}.

\begin{figure}[tbp]
\includegraphics[width=1\linewidth, draft=false]{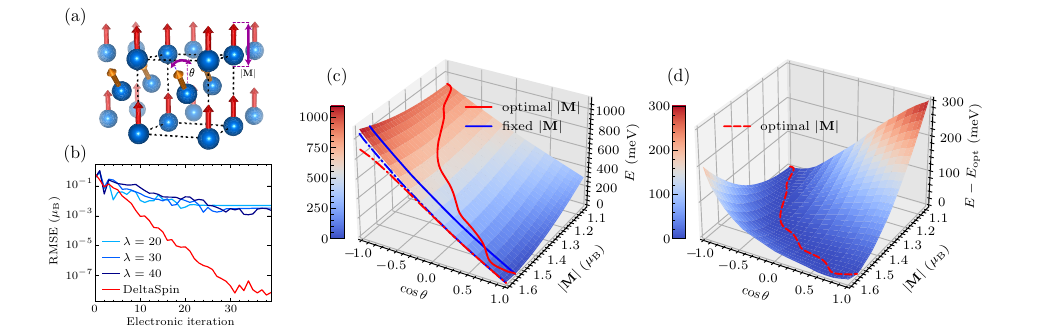}
\caption{\label{fig:iron} Potential energy surface (PES) study of bcc-Fe with magnetic excitation.
(a) Structure of bcc-Fe and spin configuration with deviated included angle of two neighboring \ce{Fe} magnetic moments $\theta = \langle\mathbf{M}_{\mathrm{Fe_1}}, \mathbf{M}_{\mathrm{Fe_2}}\rangle$ and magnitude $|\mathbf{M}|$. 
(b) Comparison of convergence between \textit{DeltaSpin} and former quadratic cDFT method with different $\lambda$. X axis: number of electronic iterations. Y axis: root-mean-square error (RMSE) between the obtained and the desired magnetization. 
(c) PES as a function of $\theta$ and $|\mathbf{M}|$. Solid and dashed lines respectively denote single-variable PES and its projection. Blue line indicates the PES with $|\mathbf{M}|$ fixed at ground-state value 1.54 $\mu_B$ (using radial Bessel function as $\hat{\mathcal{W}}_I$ in Eq.\,\ref{eq:ks_m}).
Red line indicates the PES with energy-minimized $|\mathbf{M}|$.
Dashed lines are their projections.
(d) PES relative to such energy-minimized path.
}
\end{figure}

Upon convergence, the magnetic effective field can be obtained efficiently based on the Hellmann–Feynman theorem as
\begin{equation}\label{eq:eff}
\mathbf{B}^{\mathrm{eff}}_I = -\fdv{L(\mathbf{M}_I^{*})}{\mathbf{M}_I^{*}} = - \bm{\lambda}_I.
\end{equation}
This suggests that by continuously updating the constraining field, $\Delta \mathbf{M}$ is minimized and an increasingly accurate estimation of the magnetic effective field is obtained, which turns out to be the constraining field $\bm{\lambda}_I$ itself. This estimation shows good correspondence with the true value, which is obtained via differentiation near the collinear limit, and a relative error of up to 4\% through the entire path of non-collinear rotation (see Fig.\,S2 in the Supplementary Material \cite{sm}). This is because Eq.\,\ref{eq:eff} is strictly correct only if $\ev{\nabla_{\mathbf{M}_{I}} \hat{\mathcal{H}}_{\mathrm{KS}}}{\phi_i}$ is negligible \cite{streib2020equation}. 

\vspace{5mm}

\noindent \textbf{Potential energy surface.} A fine-grid potential energy surface (PES) is of great significance, particularly the one considering the spin DoF. While the previous quadratic cDFT requires a predetermined multiplier $\lambda$ and may collapse during the early stage of the electronic loop (an analysis is presented in the Supplementary Material \cite{sm}), the proposed method can self-adaptively update the multiplier and exhibits a much better precision and efficiency. We studied the magnetic PES of the bcc-Fe (Fig.\,\ref{fig:iron}). The two DoFs scanned were the magnitude of atomic moments $|\mathbf{M}|$ and the included angle between the nearest neighbors' moments $\theta$. In the calculations using former quadratic constraints, the root-mean-square error (RMSE) between the obtained and desired spin configurations were stuck at approximately $10^{-2} \mu_B$ (blue lines in Fig.\,\ref{fig:iron}(b)). In comparison, the \textit{DeltaSpin} algorithm exhibited a much better convergence, where the RMSE decreased rapidly to nearly zero ($10^{-8} \mu_B$) for the same number of electronic iterations.
We found that the energy-favored magnitude increased when the spins approached the ferromagnetic (FM) order and decreased when approaching anti-ferromagnetic (AFM) order. Moreover, the magnitude-optimized energy curve was the PES calculated using the former direction-only constraining method, which is a subset of our $V(|\mathbf{M}|, \theta)$ PES. 

\begin{figure}[t]
\includegraphics[width=1\linewidth, draft=false]{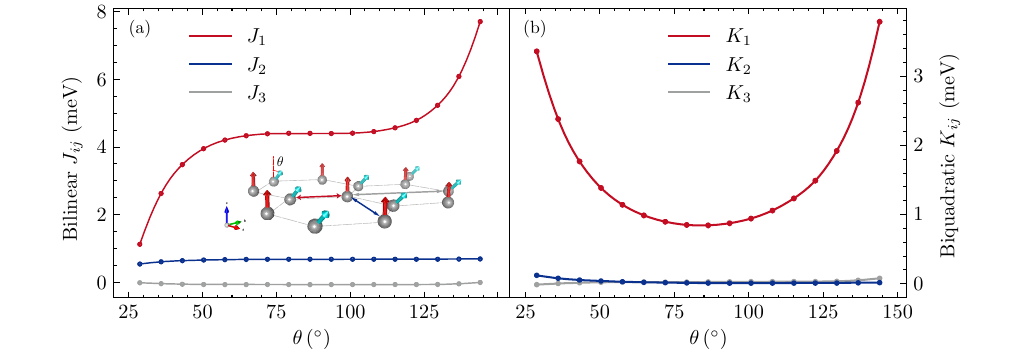}
\caption{\label{fig:jk} Configuration dependency of bilinear and biquadratic exchange interaction in monolayer \ce{CrI3}. X axis: canting angle $\theta$, that is, the included angle of two neighboring \ce{Cr} magnetic moments. Y axis: the intensity of different exchange interactions. $J_1$, $J_2$, and $J_3$ represent the interaction between the first, second, and third nearest neighbors respectively (same with $K_1$, $K_2$, $K_3$, indicated using different colors in the inset of (a)). (a) Bilinear parameter $J_{ij}$. The inset figure is an illustration of honeycomb \ce{CrI3} and the exchange interaction where Cr atoms are denoted by grey balls and O atoms are omitted. (b) Biquadratic parameter $K_{ij}$.} 
\end{figure}

\vspace{5mm}

\noindent \textbf{Effects on magnetic interactions.} One of the advantages of this method is the ability to obtain the real on-site magnetic interaction around an arbitrary magnetic configuration through simple calculations of the first derivative of energy $E$ with respect to $\mathbf{M}$. This was not possible in previous studies as achieving precision for both ${E}$ and $\delta \mathbf{M}$ was challenging. 
In Fig.\,\ref{fig:jk}, we show the calculated $J_{ij}$ and $K_{ij}$ as the bilinear exchange and biquadratic exchange parameters with variation in the magnetic configurations. The effective Hamiltonian used was $H = -\sum_{i, j > i} J_{i j} \left(\mathbf{e}_{i} \cdot \mathbf{e}_{j}\right) - \sum_{i, j > i} K_{i j} \left(\mathbf{e}_{i} \cdot \mathbf{e}_{j}\right)^2 $, where $i, j$ is the atom index. Both $J_1$ and $K_1$ exhibited strong dependency on the local magnetic configurations, as shown in Fig.\,\ref{fig:jk}. Further, $J_1$ from the previous energy-mapping strategy, which is a constant at approximately 3 meV \cite{ke2021electron}, was between the maximum and the minimum of the result obtained.


The magnetic effective fields $\fdv{E}{\mathbf{M}}$ and magnetic torques $\fdv{E}{\mathbf{e}}$ are generally considered as the indicators of magnetic interactions and could be employed as the driving forces in further dynamical simulations using real-time TDDFT or the Landau-Lifshitz-Gilbert(Bloch) method. There are two different strategies to calculate them: 
(i) Finite difference approximation directly using the definition, that is, taking derivatives of the total energy $E$ with respect to $\mathbf{M}_I^{*}$; 
(ii) Hellmann–Feynman approximation, that is, the constraining field, explained above.

We calculated such constraining-field-approximated magnetic torques and compared them to the finite-differentiated ones in bcc-Fe and monolayer \ce{CrI3} (see Fig.\,S2). The maximum relative error was about 4\%, which was attributed to the non-collinear contribution in Kohn-Sham functional $\ev{\nabla_{\mathbf{M}_{I}} \hat{\mathcal{H}}_{\mathrm{KS}}}{\phi_i}$, and perfectly acceptable in some cases.
Notice that the difference was negligible in the collinear limit, that is, in the neighborhood of FM and AFM order, both in Fe and in \ce{CrI3}. It means that Hellmann–Feynman approximation is a efficient and reliable choice given that the complexity of explicit differentiation is at least $\mathcal{O}(N)$ ($N$ is the number of atoms) whereas to obtain constraining field only one calculation is needed. 
In addition, if a delicate description instead of a crude estimate of magnetic dynamics is needed, one can always apply an infinitesimal change to spin moments, recalculate the total energy, and take the derivatives. This procedure is also unachievable for $\lambda$-fixed constraining formalism wherein the inadequate precision of energy leads to poor precision of magnetic torques.

\begin{figure}[tb]
\includegraphics[width=1\linewidth, draft=false]{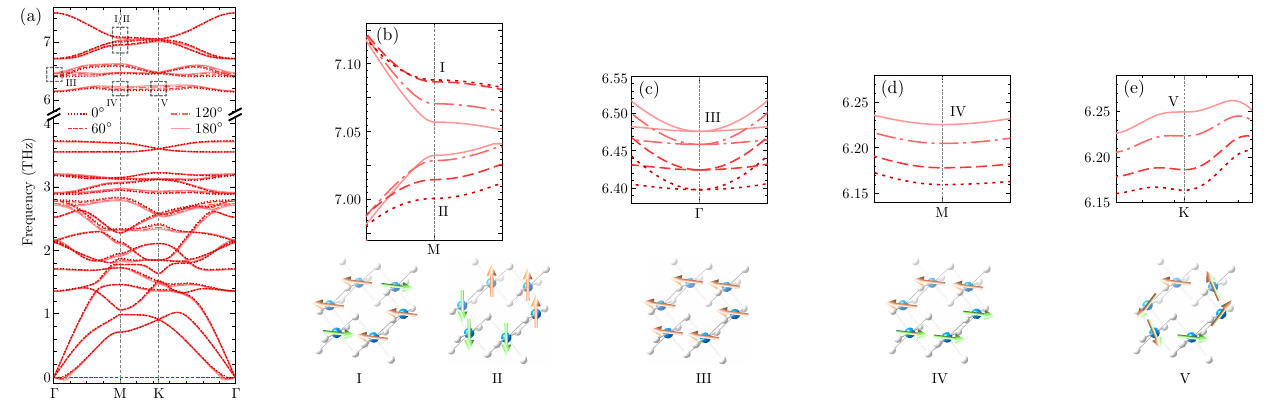}
\caption{\label{fig:pband} Phonon structure of monolayer \ce{CrI3}'s different magnetic states. (a) Overall phonon spectrum at $\theta=0^\circ, 60^\circ, 120^\circ, 180^\circ$. Dashed blocks indicate five modes (indexed by roman numerals) with significant frequency change owing to magnetic excitation. Physically-irreducible representations: 
I: $\mathrm{M_1^-}$;
II: $\mathrm{M_1^+}$;
III: $\mathrm{\Gamma_2^-+\Gamma_3^-}$;
IV: $\mathrm{M_1^+}$; 
V: $\mathrm{K_1}$. 
(b-e) Zoom image of mode I, II, III, IV, and V with the schematic atomic displacement on the bottom side. Azure and white balls denote Cr and I atoms, respectively. The arrows indicate the instant direction of their motion, where the green and orange ones represent two opposite directions.}
\end{figure}

\vspace{5mm}

\noindent \textbf{Effects on lattice dynamics.} The variation in spin configuration has a non-trivial influence on phonon behaviors, and this study proposed a straightforward method to identify such couplings.
This was performed by combining the frozen phonon method \cite{phonopy} with \textit{DeltaSpin}. In particular, the system's energy must be precisely calculated at the atomic and magnetic excitations. Simultaneously, the magnetic configuration must be prevented from ``drifting away". \textit{DeltaSpin} can limit the energy and on-site moment error to $10^{-9}\,\mathrm{eV}$ and $10^{-7}\,\mu_B$, respectively, while maintaining efficiency with more than one hundred atoms; to the best of our knowledge, it is the only method that can access this type of phonon calculations. 
Four selected spin-fluctuating states were obtained for \ce{CrI3} (Fig.\,\ref{fig:pband}). Only the moments of chromium were constrained, while iodine atoms were fully relaxed, owing to the functionality of \textit{DeltaSpin} to selectively constrain atoms or components. We found five modes that were significantly influenced by magnetic excitation, all of which appeared in the high-frequency branches where the Cr vibrations dominated (Fig.\,\ref{fig:pband}(a)). Regardless of spin-orbit coupling and anti-symmetric exchange, these frequency shifts can be roughly explained by the change in the ``Heisenberg-only” force constant. 
$\left[\pdv*{J_{ij}}{\mathbf{r}}\right]\left[\mathbf{e}_{i} \cdot \mathbf{e}_{j}\right]$. 
The second term depends on the spin configuration explicitly. The first term, which is mainly contributed by the competition between the AFM $t_{2g}$-$t_{2g}$ and FM $t_{2g}$-$e_g$ interactions \cite{bsprb2022}, also experiences considerable changes because of the different occupancies of $e_g$ and $t_{2g}$ orbitals across all four configurations.
The existence of ``collective-motion" modes, in which the distance between any two Cr atoms remains unchanged (Fig.\,\ref{fig:pband}(c), $\mathrm{\Gamma_2^-+\Gamma_3^-}$), indicates that the interaction between Cr and I atoms, that is, the metal-ligand interaction, is also strongly affected by only changing the on-site moments of metal atoms. Moreover, the moments of I relax from $10^{-1} \mu_B$ to approximately zero as $\theta$ increases from $0^\circ$ to $180^\circ$, whose importance has been demonstrated by previous research \cite{yeprb2014, yeprb2015, Logemann_JPCM_2017, Solovyev_PRB_2021}. Using this algorithm, we can obtain the spin-lattice interaction information. Consequently, the process starts from the objective of obtaining a particular magnetic configuration to determine a feasible method of selective excitation via lattice vibrations. Thus, this can guide ultrafast THz experiments through resonant excitation of infrared (IR) \cite{foteinopoulou2019phonon} or Raman-active phonons \cite{forst2015mode}.

\begin{figure}[t]
\includegraphics[width=1\linewidth, draft=false]{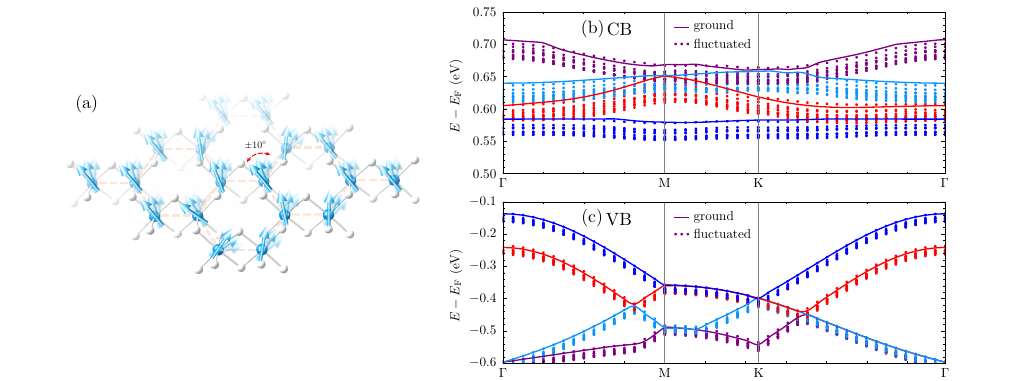}
\caption{\label{fig:eband_fluctuate} Band dispersion near conduction band minimum (CBM) and valence band maximum (VBM), resulting from randomly fluctuated spin configurations. (a) Schematic of monolayer \ce{CrI3} with randomly fluctuated spins. (b-c) Band dispersion for several selected branches. Solid curves represent ground-state bands while scatters represent corresponding fluctuated ones. Different colors are used to indicate different branches. 
}
\end{figure}

\vspace{5mm}

\noindent \textbf{Effects on electronic structure.} Spin fluctuations have a prominent effect on the electronic structure such as the band structure and orbital character of the Fermi surface \cite{maier2011d, fanfarillo2016orbital, rhodes2018scaling}. First-principles calculations of electronic structures with spin fluctuations are a critical input for further evaluating the many-body effect \cite{held2006realistic}. Here, we demonstrate that the band structure is profoundly affected by spin fluctuations using the constrained method (Fig.\,\ref{fig:eband_fluctuate}). We modeled several spin-fluctuating configurations near the ground state of monolayer \ce{CrI3}, wherein magnetic moments were drawn at random from the uniform distribution within the FM order $\pm 10^\circ$. Their ``excited" Kohn-Sham orbitals were then precisely calculated following the DFT routine, which distinguished our method from TDDFT utilizing the expansion of the ground state. Different branches from the obtained bands showed slightly different broadening behaviors with amplitudes varying in the range of 30-50 meV; consequently, the bandgap shifted. In addition, we obtained several electronic states with different canting angles $\theta$ in the transition from the FM to the Néel AFM phase of \ce{CrI3}. These exhibited prominent variation in the band gap, Fermi surface, and topological features (see Fig.\,S4-5 in the Supplementary Material \cite{sm}). Thus, this method can be a potentially useful strategy to access a spin-renormalized band structure that fully considers the coupling between spin and electronic DoF \cite{Kuwabara_PRL_2000}, which is exceptionally useful for investigating fundamental low-energy physics or exploring rich functionalities in dynamic or optical properties.

\begin{figure}[tb]
\includegraphics[width=1\linewidth, draft=false]{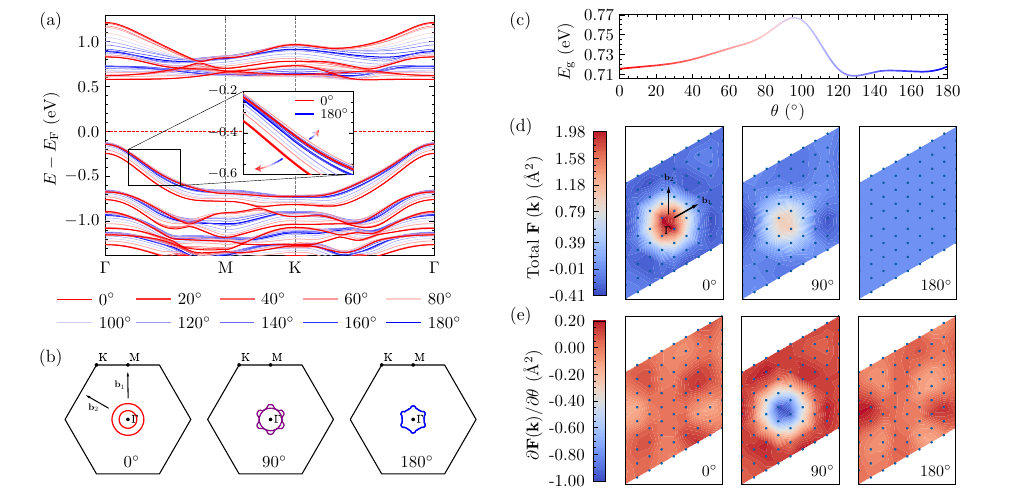}
\caption{\label{fig:eband} Electronic structure of monolayer \ce{CrI3} of different canting angles $\theta$. (a) Band structure near Fermi level. Ten band structures are in the same figure, aligned according to their Fermi levels. The inset shows the two-fold splitting of VBM from FM phase (red, $\theta = 0^\circ$) to Néel AFM phase (blue, $\theta = 180^\circ$). 
(b) Fermi surface in two dimensions with $E = E_{\mathrm{F}} - 0.31 \mathrm{eV} $. 
(c) Band gap as a function of $\theta$. 
(d) Total Berry curvature and its gradient with respect to $\theta$ as a function of $\mathbf{k}$ in reciprocal primitive cell at $\theta=0^\circ, 90^\circ, 180^\circ$.
(e) The derivative of Berry curvature with respect to $\theta$ as a function of $\mathbf{k}$ in reciprocal primitive cell at $\theta=0^\circ, 90^\circ, 180^\circ$.
}
\end{figure}

The wave function is solved precisely via DeltaSpin for spin-fluctuating states. Thus, we are able to depict more detailed electronic structure.
Strong interplay between magnetism and topology of electronic states was found, bringing forth rich functionalities in dynamical or optical properties. The topological sensibility to the magnetic configuration was observed for the first time using cDFT. We modeled ten excited states with different canting angles $\theta$ in the transition from FM to Néel AFM phase of monolayer \ce{CrI3}. They exhibited significant variation in band structure, band gap, and Fermi surface as shown in Fig.\,\ref{fig:eband}. 
We noticed the splitting of every two-fold degenerate band when $\theta$ went from $180^\circ$ to $0^\circ$, following the parity-time (PT) symmetry breaking from AFM to FM phase. It could be further demonstrated in Fermi surface, where one hexagon-shape band in AFM phase split into two concentric circle-shaped valence bands in FM phase (Fig.\,\ref{fig:eband}(b)). This process happened on $\mathrm{\Gamma}$-M path first, then on $\mathrm{\Gamma}$-K. Simultaneously, the band gap went through a variation of about $50$ meV.
Moreover, we demonstrated the change in Berry curvature $\mathbf{F}(\mathbf{k})$ and its sensitivity $\partial\mathbf{F}(\mathbf{k})/\partial\theta$ throughout the process (see Fig.\,\ref{fig:eband}(d-e)), These were calculated using Fukui's formalism, defined as $\mathbf{F}(\mathbf{k}) \equiv \ln\left[ U_1\left(\mathbf{k}\right) U_2\left(\mathbf{k}+\hat{1}\right) U_1\left(\mathbf{k}+\hat{2}\right)^{-1} U_2\left(\mathbf{k}\right)^{-1}\right]$, where $\hat{1}(\hat{2})$ denotes the reciprocal basis vector and $U_{1(2)}$ represents a purposely defined U(1) link variable \cite{fukui2005chern}. However, the Chern number \cite{li2022chern}, obtained by integrating the Berry curvature over the Brillouin zone, shows no dependence on the magnetic configuration because \ce{CrI3} is, as a whole, a trivial insulator.

\vspace{5mm}

\noindent {\large\textbf{Discussion}}

\noindent This study proposes a self-adaptive spin-constrained DFT method wherein the spin fluctuations or magnetic excitation were incorporated along with spin-orbit coupling and non-collinear magnetic configurations, capable of dealing with full degrees of freedom of the amplitude and rotation angle from both magnetic and ligand atoms \cite{Solovyev_PRB_2021}. This enabled us to gain further insight into spin fluctuations from the electron, lattice, and magnetic perspectives, and also get access to the wave function of magnetic excited states. Here we demonstrate the broading of band because of the random perturbations of magnetic moments, while this can be easily scaled to different temperature and the broading with $k$ dependence can be an indicator to real coupling between spin and other DoFs such as electron and lattice. In addition, the obtained precise atomic forces and magnetic effective fields could be employed as the driving forces in further dynamical simulations such as Landau–Lifshitz–Gilbert (or Landau-Lifshitz-Bloch) or TDDFT \cite{garanin1997fokker, gilbert2004phenomenological}. Moreover, the rapid calculation of energy and its corresponding derivatives indicate its ability as a systematic data generator for machine learning surrogate models \cite{novikov2020machine, Tranchida_JCP, Tranchida_npj, eckhoff2021high, yu2022complex} and is expected to be revolutionary with its ability to obtain arbitrary spin-lattice configurations with first-principles precision and high efficiency.

\vspace{5mm}

\noindent {\large\textbf{Methods}}

\noindent \textbf{Spin-constrained approach.} The spin-constrained DFT calculations have been performed using the self-adaptive spin-constraining algorithm \textit{DeltaSpin}, which has been implemented as a loadable module for Vienna Ab initio Simulation Package (VASP) \cite{kresse1994ab,kresse1996efficiency, kresse1996efficient}. The tolerance $\varepsilon_\mathbf{M}$ for all the constrained local magnetic moments was $10^{-8}\mu_B$. In Fe bulk, the magnetic moments of all Fe atoms were constrained to required values. In \ce{CrI3}, only the magnetic moments of Cr atoms were constrained and those of I atoms were set to be fully relaxed. For the finite difference calculations of magnetic effective fields, a $10^{-3}$\,rad interval was used to approximate an infinitesimal variation when calculating $\Delta E/\Delta \mathbf{e}$.

\vspace{5mm}

\noindent \textbf{Density functional theory calculations.} All the structural optimization and
electronic structure calculations have been performed with VASP at the level of density functional theory with the Perdew-Burke-Ernzerhof (PBE) functional. The energy cut off was 600\,eV for the plane-wave basis of the valence electrons in bulk Fe and 520\,eV for monolayer \ce{CrI3}. The DFT+U approach with $U = 5.3\,\mathrm{eV}$ is applied for Cr in \ce{CrI3}. 
Total energy tolerance for electronic structure minimization was $10^{-8}$\,eV and a $\Gamma$-centered mesh with k-spacing $0.06\pi\,\text{\AA}{}^{-1}$ was applied for both two systems. 
The VASPBERRY code \cite{Kim_VASPBERRY_2018} was used to calculate Berry curvature and Chern number directly from the output of VASP. 
For the phonon calculations, Phonopy was used to create a $3 \times 3 \times 1$ supercell structure for monolayer \ce{CrI3} and VASP was then employed to calculate the force constants. 

\vspace{5mm}

\noindent \textbf{Local mapping of the spin Hamiltonian.} Inspired by the energy-mapping method \cite{gordon2021magnetic}, we defined ``canting angle" $\theta$ in honeycomb-like \ce{MX3} as Fig.\,S3. We fit a ``local'' second-order polynomial around $\theta^*$ to the total energy of certain configurations with different $\theta$ as follows:
\begin{align}
    H&=-\sum_{i < j} J_{i j} \left(\hat{\mathbf{e}}_{i} \cdot \hat{\mathbf{e}}_{j}\right) - \sum_{i < j} K_{i j} \left(\hat{\mathbf{e}}_{i} \cdot \hat{\mathbf{e}}_{j}\right)^2 \\
    &= a_0(\theta^*) + a_1(\theta^*) \cdot \cos{\theta} + a_2(\theta^*) \cdot \cos^2{\theta},
\end{align}
where $\theta$ is in the neighborhood of $\theta^*$, that is, $\theta \in (\theta^* - \delta \theta, \theta^* +  \delta \theta)$. Notice that all coefficients of the polynomial implicitly depends on the centered $\theta^*$. After deduction, $\theta^*$-centered pair-wise exchange parameters from the first nearest neighbor (1-NN) up to 3-NN can be solved via
\begin{equation}
\left(\begin{array}{ccc}
6 & 0 & 6 \\
2 & 8 & 6 \\
4 & 8 & 0
\end{array}\right) \left(\begin{array}{c}
J_{1}, K_{1} \\
J_{2}, K_{2} \\
J_{3}, K_{3}
\end{array}\right)
= 
-\left(\begin{array}{c}
a_1^\mathrm{N}, a_2^\mathrm{N} \\
a_1^{\mathrm{Z}}, a_2^{\mathrm{Z}} \\
a_1^{\mathrm{S}}, a_2^{\mathrm{S}}
\end{array}\right),
\end{equation}
where the dependency of all quantities on $\theta^*$ are omitted. N for Néel, Z for Zigzag, S for Stripy. These exchange parameters can be treated as the local exchange parameters at $\theta^*$.

\vspace{10mm}

\vspace{5mm}
\noindent \textbf{Acknowledgements}\\
\noindent We appreciate Han Wang, Kun Cao, C. Freysoldt for their helpful discussions.

\vspace{5mm}
\noindent \textbf{Funding}\\
\noindent This work was funded by the National Natural Science Foundation of China (Grant Nos. 51790494 and 12088101).

\vspace{5mm}
\noindent \textbf{Availability of data and materials}\\
\noindent All the relevant data discussed in the present paper are available from the authors on request. Spin-constrained calculations were conducted with the \textit{DeltaSpin} code (https://github.com/\\
caizefeng/DeltaSpin). 

\vspace{10mm}
\noindent {\large\textbf{Declarations}}

\vspace{2mm}
\noindent \textbf{Ethics approval and consent to participate}\\
\noindent Not Applicable.

\vspace{5mm}
\noindent \textbf{Consent for publication}\\
\noindent Not Applicable.

\vspace{5mm}
\noindent \textbf{Competing interests}\\
\noindent YX is an editorial board member for Quantum Frontiers and was not involved
in the editorial review, or the decision to publish, this article. All authors declare
that there are no competing interests.

\vspace{5mm}
\noindent \textbf{Author contributions}\\
\noindent BX conceived and instructed the research. Under the guidance of BX, ZC performed the theoretical derivation, code implementation, and data analysis. KW, YX, and SW contributed to the interpretation of the results. ZC and BX wrote the manuscript with the input from KW, YX, and SW. All authors read and approved the final manuscript.




\bibliography{main}




\end{document}


\renewcommand{\thetable}{S\arabic{table}}
\renewcommand{\thefigure}{S\arabic{figure}}


\title{Supplementary Material: A Self-Adaptive First-Principles Approach for Magnetic Excited States}

\author{Zefeng Cai}
\affiliation{%
 School of Materials Science and Engineering, Tsinghua University, Beijing 100084, People’s Republic of China}%

\author{Ke Wang}
  \email{wang-ke@mail.tsinghua.edu.cn}
\affiliation{%
 School of Materials Science and Engineering, Tsinghua University, Beijing 100084, People’s Republic of China}%


\author{Yong Xu}
\affiliation{
 Department of Physics, Tsinghua University, Beijing 100084, People’s Republic of China
}%

\author{Su-Huai Wei}
\affiliation{
 Beijing Computational Science Research Center, Beijing 100193, People’s Republic of China
}%

\author{Ben Xu}
 \email{bxu@gscaep.ac.cn}
\affiliation{
 Graduate School of China Academy of Engineering Physics, Beijing 100088, People’s Republic of China
}%

\date{\today}


\maketitle


\subsection{Analysis of $\lambda$-fixed spin constraining methods and comparison with the self-adaptive spin-constrained method (DeltaSpin)}

In previous cDFT \cite{ma2015constrained} method, a fixed multiplier $\lambda$ is introduced into the Lagrangian of the system, which takes the form of
\begin{align}
    L\left(\left\{\phi_i\right\}, \lambda, \left\{\eta_i\right\}\right)&=
    E_{\mathrm{KS}}\left[\left\{\phi_i\right\}\right] -\sum_i \eta_i\left[\bra{\phi_i}\ket{\phi_j}-\delta_{ij}\right] \nonumber\\
    &+\lambda\sum_I \left[\mathbf{M}_I\left(\left\{\phi_i\right\}\right)-\mathbf{M}_I^*\right]^2.
\end{align}
The constraining potential, which can be obtained by taking variational derivatives with respect to orbitals $\phi_i$, carries the pre-factor of $2\lambda\sum_{I}(\mathbf{M}_I - \mathbf{M}_I^{*})$. The optimization of this Lagrangian function is performed merely by electronic minimization, that is, self-consistently solving $\phi_i$, where $\lambda$ is set as a constant. 

Several factors can sabotage this process. Firstly, the fixed $\lambda$ makes the Hamiltonian unstable to be diagonalized particularly when the initial moments are far from the target (large $2\lambda\sum_{I}(\mathbf{M}_I - \mathbf{M}_I^{*})$), and ending up with rough precision. Although one can achieve fair convergence by setting small $\lambda$ at an early stage and gradually increasing it, the process is purely empirical and oversimplified. Secondly, every component of every magnetic moment does not share the same $\lambda$ in the real cases of spin fluctuations. Each magnetic component exhibits different deviation from the original ground state, which obviously requires that the constraining field $\lambda$ depends on the local magnetic environment and varies from component to component. Furthermore, the Jacobian matrix of constraint condition has all-zero entries ($2\left(\mathbf{M}_I - \mathbf{M}_I^{*}\right)\pdv{\mathbf{M}_I}{\phi_i}$) when every $\mathbf{M}_I = \mathbf{M}_I^{*}$, which makes the quadratic penalty a theoretically, to be more specific, rank-wisely illegal constraint for determining the target minimum via Lagrangian multiplier formalism.

The following tests proved that the total energy obtained using DeltaSpin can be viewed as the limit of those using $\lambda$-fixed methods as $\lambda$ tends to $+\infty$, which is inaccessible owing to numerical instability of $\lambda$-fixed methods (see Tab.\,\ref{tab:compare}).

\begin{table}[htbp]
    \centering
    \caption{Total energy of a non-ground state constrained using VASP and DeltaSpin.}
    \begin{threeparttable}
    \begin{tabular}{wc{2.1cm}wc{1.1cm}wc{3.5cm}wc{2.6cm}}
    \toprule
    \toprule
        Method  & $\lambda$ & E\tnote{a} &  RMSE\tnote{b} \\
         \cmidrule(lr){3-3} \cmidrule(lr){4-4}
        &  & (\si{eV}) & ($\mu_{\mathrm{B}}$)\\
        VASP & 10 & -16.91540089 & \num{3.98E-3} \\
        & 30 & -16.91496141 & \num{1.36E-4}  \\
         & 40 & -16.91490396 & \num{1.02E-4} \\
         & 45 & -16.91488467 & \num{0.91E-4} \\
         & 50\tnote{c} & -- & -- \\
        DeltaSpin & -- & -16.91472775 & \num{1.18E-8} \\
         \bottomrule
         \bottomrule
    \end{tabular}
    \begin{tablenotes}
      \footnotesize 
      \item[a] The energy from VASP has already been subtracted by $E_{\text{penalty}} = \lambda\sum_I \left[\mathbf{M}_I\left(\left\{\phi_i\right\}\right)-\mathbf{M}_I^*\right]^2$.
      \item[b] Root mean square error between obtained and target on-site magnetic moments. 
      \item[c] Diagonalization crashed due to excessive multiplier $\lambda$.
    \end{tablenotes}
    \end{threeparttable}
    \label{tab:compare}
\end{table}

\subsection{Comparison of constraint-free spin-fluctuating states obtained with and without Lagrangian-based constraints}

Certain non-collinear spin-fluctuating states can be obtained using first-principles approaches without imposing explicit constraints. Spin spiral, for example, may be conveniently modeled using a generalization of the Bloch condition. While the size of the system, that is, the number of atoms or the volume of the cell, can be the same as the ground state if using this means, the appropriate cutoff energy of the basis set may become considerably larger.
Alternatively, we can also model spin spiral as a non-collinear magnetic configuration with a unusual periodicity along propagation vector, and then conduct a self-consistent DFT routine while maintaining such configuration, which can be easily done using DeltaSpin.

We obtained several spin spirals in FCC iron using these two different approaches. The propagation vector $\mathbf{q}$ pointed in the direction of \hkl(001) with $\left|\mathbf{q}\right|$ ranging from 0 to 1. The energy-minimized magnitude of the spins decreased remarkably as $\left|\mathbf{q}\right|$ went from 0 to 1. Using DeltaSpin, we calculated the total energy while constraining the spin moments at their energy-favored magnitude (red curve), which exhibited a good correspondence with that from the constraint-free approach (blue curve), and at a fixed 1.7805\,${\mu}_B$ (orange dash), which was its magnitude at $\left|\mathbf{q}\right| = 0$, that is, FM order.
These results demonstrate the accuracy and versatility of self-adaptive spin-constrained method DeltaSpin.

\begin{figure}[tb]
\includegraphics[width=0.8\linewidth, draft=false]{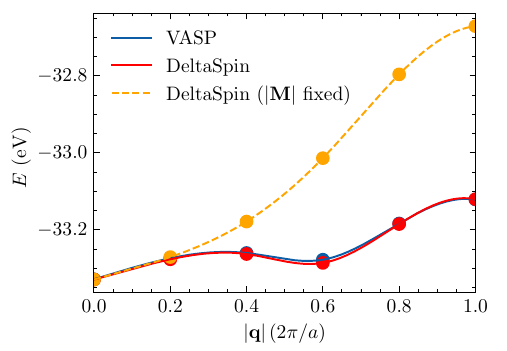}
\caption{\label{fig:spiral} Total energy of spin spiral in FCC iron calculated using generalized Bloch condition and DeltaSpin, respectively. $\mathbf{q}$ points in the direction of \hkl(001). $\left|\mathbf{q}\right|$ ranges from 0 to 1 with $2\pi/a$ as the unit. $a$ denotes the lattice constant of FCC iron.
}
\end{figure}

\subsection{Two different strategies to calculate magnetic torques using DeltaSpin}

There are two different strategies to calculate magnetic effective fields $\fdv{E}{\mathbf{M}}$ and magnetic torques $\fdv{E}{\mathbf{e}}$, which could be employed as the driving forces in further dynamical simulations using real-time TDDFT or the Landau-Lifshitz-Gilbert(Bloch) method.
(i) Finite difference approximation directly using the definition, that is, taking derivatives of the total energy $E$ with respect to $\mathbf{M}_I^{*}$; 
(ii) Hellmann–Feynman approximation, explained in the main text.

\begin{figure}[tb]
\includegraphics[width=0.8\linewidth, draft=false]{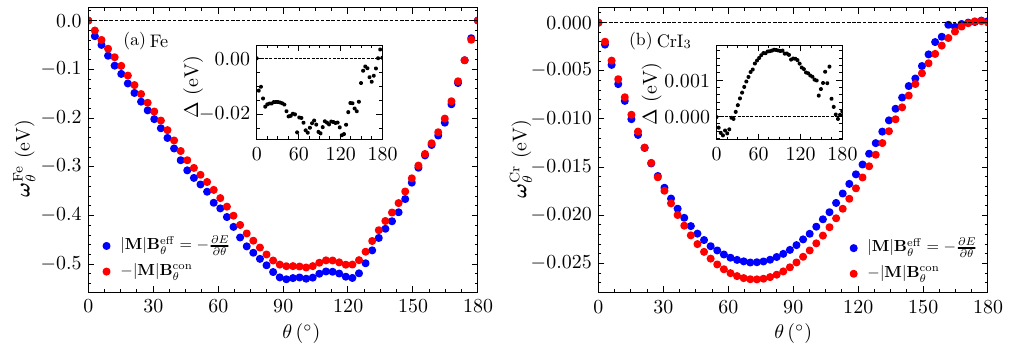}
\caption{\label{fig:eff} Comparison of the magnetic torques calculated using finite difference $-\frac{\partial E}{\partial \theta}$ (blue scatters) and Hellmann–Feynman approximation $-|\mathbf{M}|\mathbf{B}^{\mathrm{con}}_{\theta}$ (red scatters), respectively. The X axis represents the canting angle $\theta$ from ferromagnetic (FM) order ($\theta = 0^\circ$) to anti-ferromagnetic (AFM) order ($\theta = 180^\circ$). Insets show the difference between them. (a) BCC \ce{Fe} (b) Monolayer \ce{CrI3}. Here for \ce{CrI3}, $\theta = 180^\circ$ corresponds to Néel phase, one of a few AFM orders it has.}
\end{figure}

We compared such constraining-field-approximated magnetic torques to the finite-differentiated ones in BCC iron and monolayer \ce{CrI3} (see Fig.\,\ref{fig:eff}). The maximum relative error was about 4\%, which was attributed to the non-collinear contribution in Kohn-Sham functional $\ev{\nabla_{\mathbf{M}_{I}} \hat{\mathcal{H}}_{\mathrm{KS}}}{\phi_i}$, and perfectly acceptable in some cases.
Notice that the difference was negligible in the collinear limit, that is, in the neighborhood of FM and AFM order, both in Fe and in \ce{CrI3}. 

Hellmann–Feynman approximation is a efficient and reliable choice given that the complexity of explicit differentiation is at least $\mathcal{O}(N)$ ($N$ is the number of atoms) whereas to obtain constraining field only one calculation is needed. 
In addition, if a delicate description instead of a crude estimate of magnetic dynamics is needed, one can always apply an infinitesimal change to spin moments, recalculate the total energy, and take the derivatives. This procedure is also unachievable for $\lambda$-fixed constraining formalism wherein the inadequate precision of energy leads to poor precision of magnetic torques. 

\subsection{More details of magnetic interaction parameter fitting}

\begin{figure}[b]
\centering
\includegraphics[width=0.8\textwidth]{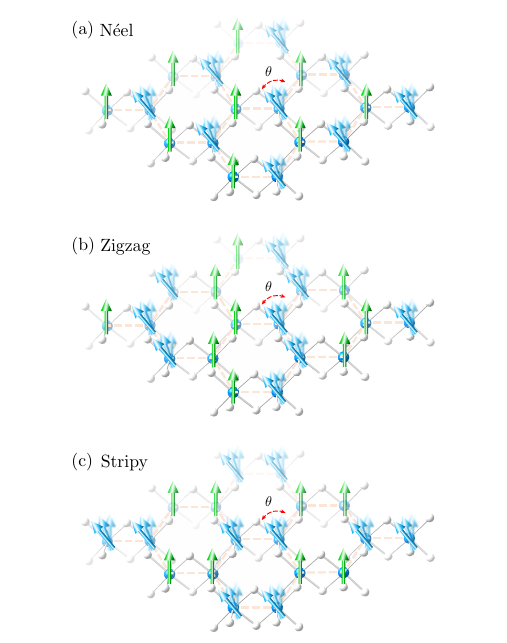}
\caption{Schematic of generalized Néel, Zigzag, Stripy AFM order in monolayer \ce{MX3}. Blue spins cant at angle $\theta$ whereas green ones remain still. (a) Néel. (b) Zigzag. (c) Stripy.}
\label{sfig:jk_scheme}
\end{figure}

Inspired by the energy-mapping method \cite{gordon2021magnetic}, we defined ``canting angle" $\theta$ in honeycomb-like \ce{MX3} as Fig.\,\ref{sfig:jk_scheme}. We fit a ``local'' second-order polynomial around $\theta^*$ to the total energy of certain configurations with different $\theta$ as follows:
\begin{align}
    H&=-\sum_{i < j} J_{i j} \left(\hat{\mathbf{e}}_{i} \cdot \hat{\mathbf{e}}_{j}\right) - \sum_{i < j} K_{i j} \left(\hat{\mathbf{e}}_{i} \cdot \hat{\mathbf{e}}_{j}\right)^2 \\
    &= a_0(\theta^*) + a_1(\theta^*) \cdot \cos{\theta} + a_2(\theta^*) \cdot \cos^2{\theta},
\end{align}
where $\theta$ is in the neighborhood of $\theta^*$, that is, $\theta \in (\theta^* - \delta \theta, \theta^* +  \delta \theta)$. Notice that all coefficients of the polynomial implicitly depends on the centered $\theta^*$. 

After deduction, $\theta^*$-centered pair-wise exchange parameters from the first nearest neighbor (1-NN) up to 3-NN can be solved via
\begin{equation}
\left(\begin{array}{ccc}
6 & 0 & 6 \\
2 & 8 & 6 \\
4 & 8 & 0
\end{array}\right) \left(\begin{array}{c}
J_{1}, K_{1} \\
J_{2}, K_{2} \\
J_{3}, K_{3}
\end{array}\right)
= 
-\left(\begin{array}{c}
a_1^\mathrm{N}, a_2^\mathrm{N} \\
a_1^{\mathrm{Z}}, a_2^{\mathrm{Z}} \\
a_1^{\mathrm{S}}, a_2^{\mathrm{S}}
\end{array}\right),
\end{equation}
where the dependency of all quantities on $\theta^*$ are omitted. N for Néel, Z for Zigzag, S for Stripy. These exchange parameters can be treated as the local exchange parameters at $\theta^*$.

\bibliography{main}